\documentstyle[aps,preprint]{revtex}
\title{New Type of Collective Motion for N $\sim$ Z Nuclei}
\author{Joseph N.\ Ginocchio}
\address{{\it T-5, MS B283, Theoretical Division, Los Alamos
National Laboratory\\ Los Alamos, NM 87545\\ }}
\begin{document}
\draft
\maketitle
\begin{abstract}
\noindent We study a new type of collective motion with $\alpha$
- particle type of correlations and show that it may be relevant
for N $\sim$ Z nuclei.\\

\end{abstract}
\pagebreak

New radioactive beam facilities will be studying proton rich
nuclei   with N $\sim$ Z. The limited systematic  information we
have on such nuclei with $A \ge 60$ suggests   different
collective behaviour than nuclei with N $\gg$ Z.  Although these
nuclei have large deformations, the ratio of the excitation
energy of the first angular momentum four state to the excitation
energy of the first angular momentum two   state,
$E^*_{4_1^+}/E^*_{2_1^+}$, is closer to 2.5 than 3.3, suggesting
gamma unstable collective motion. Furthermore,   the energy
spectrum for many such nuclei has $E^*_{4_1^+}
\sim E^*_{2_2^+} \sim E^*_{0_2^+}$ resembling a $\it spherical$
quadrupole vibrator.  Within the interacting boson model with
isospin and which includes   neutron-proton correlations, (IBM-3),
we have found these features exhibited in a collective motion
arising from   a dynamical symmetry of IBM-3 not discussed
heretofore. Neutron - proton correlations play a crucial   role in
this new collective motion;  for nuclei with N
$\sim$ Z we expect  neutron - proton correlations to be larger
than for nuclei with a large neutron excess because the valence
neutrons and protons are filling the  same   major shell, and
hence the radial wavefunctions of the neutrons will have a larger
spatial overlap   with the protons than neutrons filling different
major shells.

The interacting boson model has been phenomenologically
successful in describing the spectroscopy of heavy nuclei with a
large neutron   excess (IBM-2) \cite{iac87}. However, when the
valence neutrons   and protons are filling the same major shell
such  as in nuclei with
N $\sim$ Z, then isospin must be introduced \cite{ell80}, which
means a neutron-proton pair must   be included in addition to  the
proton-proton and neutron-neutron   pairs to complete the isospin
triplet.  These pairs are represented   in the IBM-3 by three
types of monopole ($s^{\dagger}_{\tau}$)   and quadrupole
($d^{\dagger}_{m,\tau}$) bosons, where $m=(-2,-1,0,1,2)$ is the
angular momentum   projection and $\tau = (1,0,-1)$ is the isospin
projection indicating   neutron, proton-neutron, and proton bosons
respectively
\cite{ell80}. To the extent that IBM-3 is valid, we expect it to
be valid for even-even nuclei; odd-odd nuclei will need
additional isoscalar bosons \cite{ell81}.

We assume the IBM-3 nuclear Hamiltonian conserves angular
momentum and isospin. The small isospin - breaking electromagnetic
interactions   can be treated in perturbation

\pagebreak

\noindent theory.  Since there are eighteen bosons (six
spatial-spin and three isospin degrees of freedom), the states of
IBM-3   transform like the symmetric representations of U(18). One
new dynamical symmetry, which includes the angular momentum
subgroup, O(3), and the isospin subgroup, SU(2), is its orthogonal
subgroup, O(18). The physical significance of O(18) is that the
eigenstates   have ``$\alpha$ particle - like'' correlations
because it leaves invariant the spin - isospin scalar two boson
(aka four nucleon)   system,

\begin{equation} {\cal I}^{\dagger} = s^{\dagger}:s^{\dagger} -
d^{\dagger}:d^{\dagger},
\label {eq_inv}
\end{equation} where : means scalar product in both angular
momentum and isospin space.

A basis for this dynamical symmetry can be determined by using
the subchain,

\begin{equation}
U(18) \supset O(18) \supset  O(15)   \times  SU_s(2)
\supset [ O(5) \supset O(3)]
\times SU_d(2) \times \ SU_s(2),
\label {eq_sub15}
\end{equation}
where $SU_s(2), SU_d(2)$ are the isospin groups for
the   monopole and quadrupole bosons respectively, and
$SU(2) = SU_s(2) + SU_d(2)$. We do not expect $O(15) \times
SU_s(2)$ to be a conserved subgroup. (We shall not discuss the
charge symmetric group chain, $U(18) \supset U(6)   \  \times
SU_c(3)$, nor the other O(18) subchain, $U(18)\supset O(18)
\supset  O(6) \times SU(2)$,   since U(6) and O(6) dynamical
symmetry have been studied extensively \cite{iac87,lev94,long}.)

The normalized basis states for this subgroup chain for $N$
bosons, representing $N$ pairs of valence nucleons, are
\begin{equation}
|N,\alpha,\delta,[T_s,T_d]^{(T)}_{T_z},
(\tau_1,\tau_2), J,M \rangle =   {\cal N}\ \left[ {\cal I^{\dagger}}
\right ]^{{N-\alpha} \over2}  |\alpha,\alpha,\delta,
[T_s,T_d]^{(T)}_{T_z}, (\tau_1,\tau_2), J,M \rangle,
\label {eq_suba}
\end{equation}
where ${\cal N} = \sqrt{\frac{(\alpha + 8)!}{2^{N
- \alpha} \left(\frac{N - \alpha}{2}
\right) ! \left( \frac{N + \alpha}{2} + 8 \right) !}}$, and
$$ |\alpha,\alpha,\delta,[T_s,T_d]^{(T)}_{T_z}, (\tau_1,\tau_2),
J,M \rangle   =
$$
\begin{equation}
\sum_p A_p [s^{\dagger}:s^{\dagger}]^{{{\alpha-\delta-T_s}\over2}
- p} \ [d^{\dagger}:d^{\dagger}]^ p
\ |\delta + T_s,\delta + T_s,\delta,[T_s,T_d]^{(T)}_{T_z},
(\tau_1,\tau_2), J,M \rangle
\label {eq_subd}
\end{equation}  where
\begin{equation} A_p = \frac { \sqrt{\left
({\alpha-\delta-T_s\over2}\right )! \ \left
({\alpha-\delta+T_s+1\over2} \right)!
\left (\delta+{13\over2}\right )!\left ({\alpha+\delta -T_s
+13\over2}\right )! \ \left ({\alpha+\delta+T_s+14\over2} \right)!
\left (T_s +{1\over2}\right )!} } { 2 ^{\alpha-\delta-T_s\over2}
\   p!\left ({\alpha-\delta-T_s\over2} - p\right )!
\ \left ({\alpha-\delta+T_s+1\over2} - p \right)! \left
(\delta+{13\over2} + p \right )!\sqrt{\left(\alpha + 7 \right
)!}}\ .
\label {eq_A}
\end{equation} Thus these states have $\alpha$-particle-like
correlations.   Although the number of monopole and quadrupole
bosons is not a conserved quantum number, the monopole and
quadrupole   isospin are conserved in this group chain. Only the
symmetric  representations of O(18) occur and hence only one
quantum number,
$\alpha$, is needed to label these representations. For $N$ bosons
the allowed values are
$\alpha = N, N-2,
\dots, \ 0$ or $1$.
$\alpha$ counts the number of bosons $\it not$ in the invariant in
(\ref{eq_inv}), and ${\cal I}$ will annihilate the states with
maximum $\alpha = N,\
{\cal I}\ |\alpha,\alpha,\delta,[T_s,T_d]^{(T)}_{T_z},
(\tau_1,\tau_2), J,M \rangle = 0$. The $O(15) \times SU_s(2)$
subgroup leaves both
$s^{\dagger}:s^{\dagger}$ and $d^{\dagger}:d^{\dagger}$
separately invariant, and
${\tilde s:\tilde s}|\dots \rangle = {\tilde d}:{\tilde d}
|\dots\rangle = 0$,\ where $|\dots\rangle = |\delta +   T_s,\delta +
T_s,\delta,[T_s,T_d]^{(T)}_{T_z}, (\tau_1,\tau_2), J,M \rangle$.
Likewise,only the symmetric representations for O(15)   occur, with
allowed eigenvalues
$\delta$ = $\alpha$, $\alpha$ - 1, \dots, 0, for each $\alpha$.
For   a given value of $\alpha$ and $\delta$, the allowed  values
of monopole isospin are $T_s = \alpha - \delta,\alpha -
\delta - 2,\dots$, 0 or 1. The allowed representations of the
$O(5) \times SU_d(2)$ subgroup are given in Table I for the
smallest values of $\delta$
\cite{pat81}, which are expected to lie lowest in energy. The
allowed values of the quadrupole isospin are $T_d =
\delta,
\delta - 1,\dots$, 0. This means that, for
$\delta$ = 0, $T = T_s$, and $\delta$ = 0 occurs only for $N - T$
even.

The  generators of the O(18) dynamical symmetry are given by   the
quadrupole operators, $P^{(2,t)} = [s^{\dagger}\tilde d]^{(2,t)} +
(-1)^t \  [d^{\dagger}{\tilde s}]^{(2,t)}$, where
$[...]^{(\ell,t)}$ means coupled to angular momentum rank $\ell$
and isospin t and $\tilde s_{\tau} =  (-1)^{\tau}s_{-\tau},\tilde
d_{m,\tau} =  (-1)^{m +\tau}d_{-m,-\tau}$, the generators of the
O(15) dynamical symmetry,
$Q^{(\ell,t)}_{M,q}= [d^{\dagger}{\tilde d}]^{(\ell,t)}_{M,q},\ell
+ t =
$ odd, and the isospin generators for the monopole bosons, ${\hat
T}_{s,q} = -\sqrt{2}\ [s^{\dagger}{\tilde s}]^{(0,1)}_{0,q}$.  The
angular   momentum generators are ${\hat J}_M = -\sqrt{30}\
Q^{(1,0)}_{M,0}$, and the total isospin is, ${\hat T} = {\hat
T}_{s}  + {\hat T}_d$, where ${\hat T}_{d,q} = -\sqrt{10}\
Q^{(0,1)}_{0,q}$ are the quadrupole boson isospin   generators.
The Casimir operator is given by
\begin{equation} C_{O(18)} =
\sum_{t=0,1,2}(-1)^t \ P^{(2,t)}:P^{(2,t)} + C_{O(15)} + {\hat
T}_s\cdot{\hat T}_s
\label {eq_cas}
\end{equation} with eigenvalues $\alpha  ( \alpha + 16)$. The
O(15) Casimir operator is given by
\begin{equation} C_{O(15)} =\sum_{\ell,t,\  \ell+t \ odd}
Q^{(\ell,t)}:Q^{(\ell,t)}
\label {eq_cas15}
\end{equation} with eigenvalues, $\delta (\delta + 13)$.

Hence an attractive interaction quadrupole interaction
will have the eigenvalue,
$$ -\kappa \langle N,\alpha,\delta, [T_s,T_d ]^{(T)}_{T_z},
(\tau_1,\tau_2),J,M | \hspace*{-0.05in} \sum_{t=0,1,2}(-1)^t
\hspace*{-0.05in}\ \  P^{(2,t)}: P^{(2,t)} | N,\alpha,\delta,
[T_s,T_d]^{(T)}_{T_z}, (\tau_1,\tau_2), J,M \ \rangle
$$
\begin{equation} = -\kappa(\alpha (\alpha + 16) - T_s(T_s+1)
-\delta (\delta +   13)).
\label {eq_p}
\end{equation} Thus, the most symmetric irreducible representation
(IR) of   O(18),
$\alpha = N$, will be the lowest in energy, whereas
the smallest   IR's of O(15), $\delta$ =0,1,etc., will be the lowest
in energy. The O(15) quantum number,
$\delta$, is similar to a phonon quantum number. However, because
of the additional monopole   and quadrupole isospin quantum
numbers, in general, there are more eigenstates than for the
usual   quadrupole vibrator which does not distinguish between
neutrons and protons. For example, for $\delta$ = 1, and $T = N$
and $T=0$, there is one $2^+$ excited state and it has $T_s = N -1$
and 1, respectively, but, for all other isospins there are two
$2^+$ excited states for each total isospin $T$ corresponding to two
different monopole isospins, $T_s = T \pm 1$. Furthermore, for
$\delta$ = 2, the $0^+$ state can only have $T_d$ =2, whereas the
$2^+,4^+$ states can have both $T_d$ = 0,2. This means that there
are a different number of states for $\delta$ = 2 depending on the
angular momentum. For example, although for
$\delta$ = 2, and $T = N$, there is only one $0^+,2^+,4^+$ excited
state and it has $T_s = N -2$, for $T = 0$, there is one $0^+$ state
which has
$T_s$ = 2, but two $2^+,4^+$ states corresponding to $T_s$ = 0
and 2, and for $T = 1$ there are two $0^+$ states corresponding to
$T_s$ = 1 and 3, but three $2^+,4^+$ states corresponding to
$T_s$ = 1, 3 and $T_d$ = 2, and $T_s$ = 1, $T_d$ = 0. On the other
hand for all other isospins, there are three
$0^+$ states ($T_s = T-2,T, T+2$), but four $2^+,4^+$ states
corresponding to $T_s = T-2,T, T+2, T_d = 2$, and $T_s = T$,
$T_d$ = 0. Most likely these states will mix in general,
separating them in energy; a detailed analysis of   nuclear
spectra with a Hamiltonian with O(18) dynamical symmetry is
underway using an IBM-3 diagonalization
\cite{sug94}. Nevertheless, we anticipate that there will be
changes in the phonon multiplet going from nuclei with $T
\ne$ 0 to nuclei with $T = 0$ because for $T
\ne$ 0 there are only 25$\%$ more $2^+,4^+$ two phonon states than
$0^+$ two phonon states, whereas for $T = 0$ there are 100$\%$
more $2^+,4^+$ two phonon states than
$0^+$ two phonon states.

Recently, the spectrum of $^{64}_{32}Ge_{32}$ has been measured
\cite{enn} and the first few excited states are shown in Table II
with O(18) and unperturbed O(15)   quantum numbers. The spectrum
indicates a phonon structure. The average measured   excitation
energy of the two phonon states $4^+_1$ and $2^+_2$,
${\bar E}^*_2 = \sum_{J_i=2^+_2,4^+_1} (2J+1)
E_{J_i}/\sum_{J_i=2^+_2,4^+_1} (2J+1) $, is 1.88 MeV and the ratio
of two phonon energy to one phonon energy is then
${\bar E}^*_{2}/E^*_{2^+_1}$ = 2.09, in between a spherical (2.0)
and deformed quadrupole O(6) vibrator   (2.5) \cite{iac87}. For $T
= 0$ nuclei, the  $2^+_1$ has
$\delta$ = 1 and $T_s$ =1 and is unique (Table I). The $4^+_1$ and
$2^+_2$ have $\delta$   =2, but there are two states each, one
with $T_s$ = 0 and one with $T_s$ = 2, which will probably mix.
Using the quadrupole interaction (\ref {eq_p}) to set the scale
of the phonon energy, the O(18) excitation energy is
$E^*_{\delta, T_s} = \kappa \ [ \delta\ (\delta + 13) +   T_s\ (T_s
+ 1)]$ and hence
${E}^*_{2,T_s}/E^*_{1,1}$ ranges between 1.88 and 2.25 for
$T_s$ = 0 or 2, respectively. An even admixture of $T_s$ = 0 and 2
for the $\delta=2$ state gives the observed ratio. Of course, a
ratio of 2.09 is also consistent with a spherical vibrator.
We can distinquish between a spherical vibrator and the O(18)
limit by the variation of the excitation energy of the first
excited state with isospin. For maximal isospin, $T = N$, the O(18)
model is equivalent to the spherical vibrator; as can be seen from
(\ref {eq_subd}) there are no correlations since $\delta + T_s =
\alpha = N$, and the number of quadrupole bosons, $N_d$, is a
conserved quantum number and $N_d= \delta$. As the isospin decreases
for a given $N, N_d$ remains a conserved quantum number for a
spherical vibrator, and hence the energy spectrum of the first
excited state remains a constant. However, for O(18), $N_d$ is not
conserved for $T
<  N$, and the expectation value, $\langle \dots |N_d | \dots
\rangle$, where
$|\dots \rangle = |N,N,\delta,[T_s,T_d]^{(T)}_{T_z},
(\tau_1,\tau_2), J,M \rangle$, is,
\begin{equation}
\langle \dots \vert N_d \vert \dots \rangle = \delta \ +
{{(N\ -\ \delta \ - \ T_s) (N \ +\  T_s\  -\  \delta \ + \ 1)} \over
{2(N
\ +\  7)}}.
\label{eq_Nd}
\end{equation}
We clearly see then for maximal isosopin, $T = N = \delta + T_s$,
$\langle \dots \vert N_d \vert \dots \rangle = \delta$, the same as
a spherical vibrator. However, for $T, T_s$ small and $N$ large,
$\langle
\dots \vert N_d \vert \dots \rangle \approx {1 \over 2} (N\ + \
1)\ + \ O(1/N)$; that is, the same for all states, and hence the
effect of the quadrupole boson energy goes to zero in the ${\it
spectrum}$, and thus the excitation energy of the first excited
state decreases.

A measurement of the B(E2) will also distinguish
between the two types of collective motion.
For nuclei with $T = 0,\  N$ even, only the isoscalar quadrupole
operator contributes to the $B(E2)$; hence, the  quadrupole operator
is $Q = Q_{sp}\sqrt3\ (1 + 2\Delta)\ P^{(2,0)}$, where $Q_{sp}$
is the single-particle quadrupole moment and $\Delta$ is the
effective charge. The B(E2) from the   ground state for O(18)
becomes
\begin{equation} B(E2:0^+_1,T=0 \rightarrow 2^+_1,T=0)_{O(18)} =
\frac{(1 +   2\Delta)^2}{3}  N(N+16)\ B(E2)_{sp} ,
\label{eq_be2}
\end{equation} The heaviest even - even nuclei with $T = 0$ for
which the $B(E2)$ are measured are $^{44}Ti$ and
$^{48}Cr$. Using no effective charge ($\Delta$ = 0), we find that
(\ref{eq_be2}) gives 0.055 and 0.138 $(eb)^2$, respectively, in
very good agreement with the   measured values 0.061 $\pm$ 0.015
and 0.133 $\pm$ 0.020
$(eb)^2$, respectively \cite{ram87}.
For spherical nuclei, $B(E2:0^+_1,T=0 \rightarrow
2^+_1,T=0)_{spherical}  = 5 (1 +   2\Delta)^2   N\ B(E2)_{sp}$\cite
{iac87}, producing a smaller $B(E2)$ with a different mass
dependence. In Fig. 1 we compare the two limits as a function of $N$.

In summary, the O(18) dynamical symmetry developed in this   paper
shows promise that it may have the correct neutron-proton
correlations to describe heavy nuclei with N $\sim$ Z. Only
additional measurements on nuclei with N $\sim$ Z will be able to
decide.

The author thanks M. Sugita for discussions. This work was
supported by the United States Department of Energy.

\pagebreak

\begin{table}
\caption{The allowed values of $(\tau_1,\tau_2)$, $T_d$, and J
for the lowest IR's of O(15) labeled by $\delta$.}
\vspace{18pt}
$$\vbox {\tabskip 2em plus 3em minus 1em \halign to \hsize{\hfil
#\hfill && #\hfil \hfil

\cr $\delta$ & $(\tau_1,\tau_2)$ & $T_d$ & J & \cr
\noalign{\hrule}
\noalign {\vskip 12pt} 0 & (0,0) & 0 & 0 \cr 1 & (1,0) & 1 & 2 \cr
2 & (2,0) & 0,2 & 2,4 \cr & (1,1) & 1 & 1,3 \cr & (0,0) & 2 & 0 \cr
3 & (3,0) & 1,3 & 0,3,4,6 \cr & (2,1) & 1,2 & 1,2,3,4,5\cr & (1,1)
& 0 & 1,3 \cr & (1,0) & 1,2,3 & 2\cr
\noalign {\vskip 12pt} }}$$

\end{table}

\begin{table}
\caption{The excitation energy $E^{*}_{J^{+}_{i}}$ in MeV of the
lowest energy levels in $^{64}Ge$ {\protect \cite{enn}} and their
O(18) and   unperturbed O(15) quantum numbers.}

\vspace{18pt}
$$\vbox {\tabskip 2em plus 3em minus 1em \halign to \hsize{\hfil
#\hfill && #\hfil \hfil

\cr $J^+_i$ & $E^*_{J^+_i}$ & $\alpha$ & $\delta$ & $T_s$
&$(\tau_1,\tau_2)$&$T_d$& \cr
\noalign{\hrule}
\noalign {\vskip 12pt}
$0^{\dagger}_1$&0.0&4&0&0&(0,0)&0 \cr
$2^{\dagger}_1$&0.902&4&1&1&(1,0)&1 \cr
$2^{\dagger}_2$&1.579&4&2&0 or 2&(2,0)&0 or 2 \cr
$4^{\dagger}_1$&2.052&4&2&0 or 2&(2,0)&0 or 2 \cr
\noalign {\vskip 12pt} }}$$

\end{table}
\pagebreak

\pagebreak
\begin{figure}
\caption
{The $B(E2)$ from the ground state to the first excited state,
relative to the single - particle value, $B(E2)_{sp}$, versus $N$ for
$T=0$ and $\Delta = 0$. The solid line is for the O(18) limit; the
dashed line is for the spherical limit.}
\end{figure}
\end{document}